\documentclass[twocolumn,showpacs,aps,prl,floatfix,amsmath,amssymb,notitlepage,superscriptaddress,longbibliography]{revtex4-1}
\usepackage{epsfig}
\usepackage{subfigure}
\usepackage{amsmath}
\usepackage{color}
\usepackage{amssymb}
\usepackage{setspace}
\usepackage{graphicx}
\usepackage{dcolumn}
\usepackage{bm}
\usepackage{times}
\usepackage{enumerate}
\usepackage{footnote}
\input epsf

\usepackage{algorithm}	
\usepackage{algpseudocode}

\newcommand{\appropto}{\mathrel{\vcenter{
  \offinterlineskip\halign{\hfil$##$\cr
    \propto\cr\noalign{\kern2pt}\sim\cr\noalign{\kern-2pt}}}}}

\graphicspath{{figures/}}

\begin{document}

\author{Per Sebastian Skardal}
\email{persebastian.skardal@trincoll.edu}
\affiliation{Department of Mathematics, Trinity College, Hartford, CT, 06106, USA}

\author{Federico Battiston}
\affiliation{Center for Networks and Complexity, Luiss University of Rome, Viale Romania 32, 00197, Rome, Italy}
\affiliation{Department of AI, Data and Decision Sciences, Luiss University of Rome, Viale Romania 32, 00197, Rome, Italy}
\affiliation{Department of Network and Data Science, Central European University, Vienna, Austria}

\author{Maxime Lucas}
\affiliation{Department of Mathematics and Namur Institute for Complex Systems (naXys), Université de Namur, Namur, Belgium}
\affiliation{Mycology Laboratory, Earth and Life Institute, Université Catholique de Louvain, Louvain-la-Neuve, Belgium}

\author{Matthew S Mizuhara}
\affiliation{Department of Mathematics and Statistics, The College of New Jersey, Ewing, NJ 08628, USA}

\author{Giovanni Petri}
\affiliation{NP Lab, Network Science Institute, Northeastern University London, London, UK}
\affiliation{Department of Physics, Northeastern University, Boston, MA 02115, USA}
\affiliation{CENTAI Institute, 10138 Torino, Italy}

\author{Yuanzhao Zhang}
\affiliation{Santa Fe Institute, Santa Fe, NM 87501, USA.}
\affiliation{Department of Physics and Astronomy, University of Rochester, Rochester, NY 14609, USA}

\title{Mixed higher-order coupling stabilizes new states}


\begin{abstract}
Understanding how higher-order interactions affect collective behavior is a central problem in nonlinear dynamics and complex systems. Most works have focused on a single higher-order coupling function, neglecting other viable choices. Here we study coupled oscillators with dyadic and three different types of higher-order couplings. By analyzing the stability of different twisted states on rings, we show that many states are stable only for certain combinations of higher-order couplings, and thus the full range of system dynamics cannot be observed unless all types of higher-order couplings are simultaneously considered.
\end{abstract}


\maketitle

Recent research on the dynamics of large, network-coupled systems has been fueled by the understanding that interactions between individual units take place not only between connected pairs, but also among larger groups of connected units~\cite{Battiston2020PhysRep,Battiston2021NatPhys,Majhi2022Interface,Yu2011JNeuro,bick2023higher,Battiston2026NatPhys}. These higher-order (or polyadic) interactions that take place between groups of three units, four units, etc. (i.e., via triadic coupling, tetradic coupling, etc.) encode coupling that is qualitatively different from typical pairwise (i.e., dyadic) interactions~\cite{Lambiotte2019NatPhys,Neuhauser2020PRE,Leon2025PhysD,stankovski2017coupling}. Moreover, in systems that model a range of phenomena including synchronization~\cite{Skardal2019PRL,Skardal2020CommPhys,Leon2022PRE,Skardal2023Chaos,Bick2023SIADS,Muolo2024ProcA,Bick2024JPhys,Costa2025CSF,Li2025PRE}, epidemic spreading and contagion~\cite{Iacopini2019NatComm,Landry2020Chaos,StOnge2021PRL}, and ecological competition~\cite{Grilli2017Nature}, higher-order interactions have been shown to alter the properties of existing dynamical states~\cite{Skardal2021PRR,Zhang2023NatComms} and give rise to phenomena not observed with dyadic interactions alone, including new dynamical states~\cite{Bick2011PRL,kundu2022higherorder}, new transitions between states~\cite{Iacopini2019NatComm,malizia2025hyperedge}, and multistability~\cite{tanaka2011multistable,Zhang2024SciAdv,Moehlis2026Chaos,WangX2026Chaos,Cai2025PRR}. Thus, higher-order interactions serve as a natural mechanism for inducing complex behaviors in coupled dynamical systems. 

Still, our understanding of precisely how higher-order interactions induce complex behaviors is incomplete.
The majority of research into the dynamics that result from higher-order interactions focuses on either a single kind of higher-order coupling~\cite{gong2019lowdimensional,komarov2015finitesizeinduced} or a single kind of higher-order coupling alongside classical dyadic coupling~\cite{lucas2020multiorder,leon2024higherordera}. In reality, however, mixed higher-order couplings often coexist. Coupled oscillator systems are an excellent example of this, as phase reduction analysis on limit-cycle oscillators first reveals dyadic interactions between oscillators as a first-order effect, then a combination of three different types of higher-order interactions that take place via either triadic or tetradic coupling~\cite{Ashwin2016PhysD,Leon2019PRE,mau2024phase}. While it has been well established that higher-order interactions give rise to complex behavior, available analytical techniques and the complexity of emerging dynamics have limited our ability to study systems with mixed types of higher-order couplings. 

In this Letter, we study the dynamics of coupled oscillators under the full set of admissible pairwise interactions and higher-order interactions that take place at second-order phase reduction~\cite{Ashwin2016PhysD,Leon2019PRE}. Focusing on ring networks and the possible twisted states that are steady-state solutions for the underlying dynamics, we find that the particular combination of the three kinds of higher-order interactions is critical in determining the stability of various twisted states~\cite{Wiley2006JNL,Wang2025PRE,Wang2026Chaos}. Importantly, while each type of higher-order interaction in isolation (or present only alongside typical dyadic interactions) stabilizes a different and complicated set of states, when the three kinds of higher-order interactions are present together a new set of stable states emerges that were not stable under any isolated type of higher-order coupling. Thus, considering the full set of higher-order interactions that are present in a model is critical to understanding the full range of collective dynamics---by including only a subset of the interactions, one may miss important dynamical states or phenomena. 

\begin{figure*}[t]
    \centering
    \includegraphics[width = 0.95\linewidth]{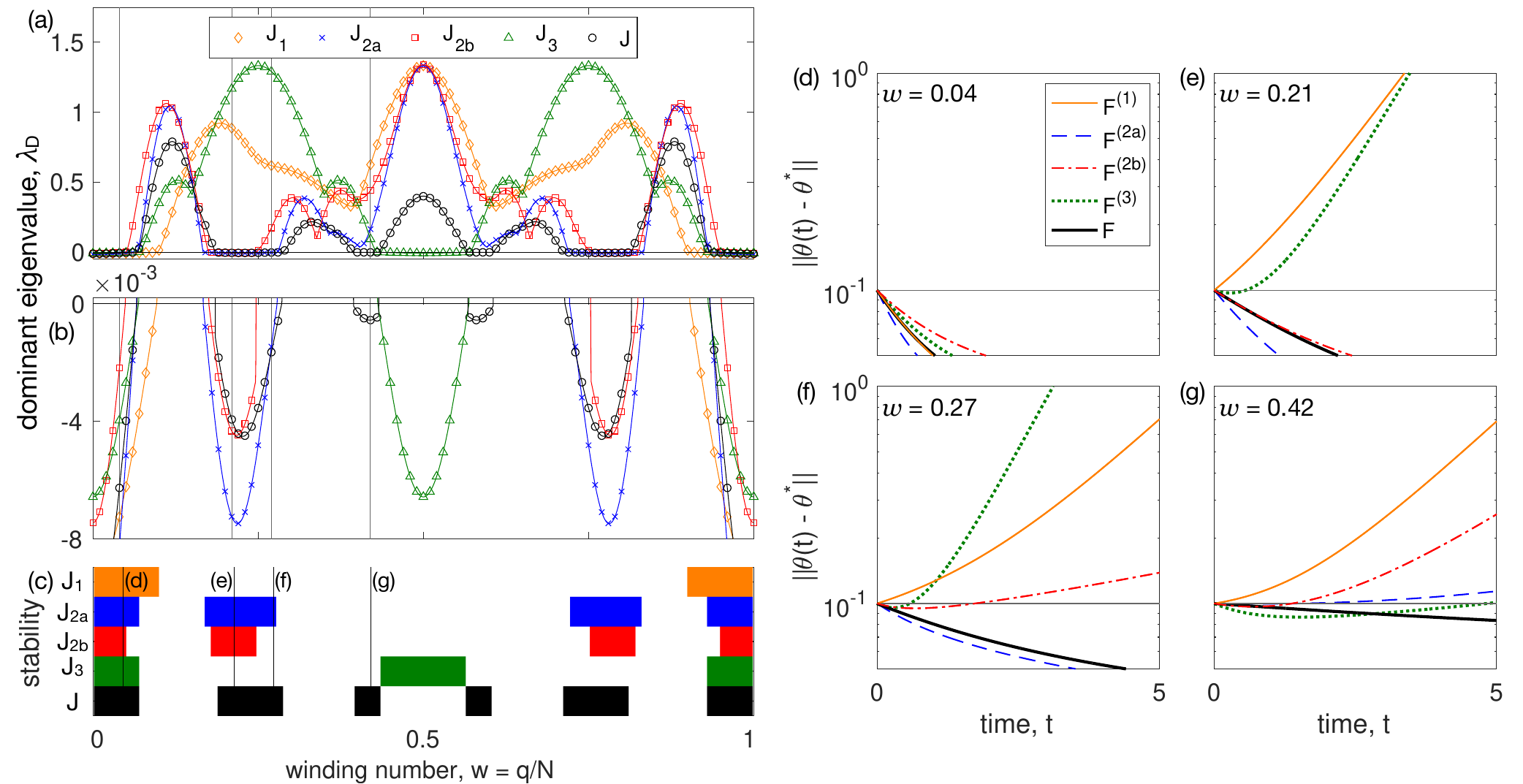}
    \caption{Stability of twisted states under isolated and mixed higher-order coupling. (a--b) For a network of size $N=100$ with interaction radius $r=3$, the dominant non-trivial eigenvalue $\lambda_D$ of the Jacobian for all possible twisted states $\bm{\theta}(w)$, where $w$ is the normalized winding number $q/N$. Results for individual Jacobians $J_1$, $J_{2a}$, $J_{2b}$, $J_3$, and the mixed coupling case $J=K_1J_1+K_{2a}J_{2a}+K_{2b}J_{2b}+K_3J_3$ are plotted in orange diamonds, blue crosses, red squares, green triangles, and black circles, respectively. Note that (b) shows a zoomed-in view of the eigenvalue near zero to more easily identify stability, i.e., when $\lambda_D<0$. We summarize the respective regions of stability for twisted states in panel (c). (d--g) The evolution of a perturbation, $\|\bm{\theta}(t)-\bm{\theta}^*\|$ towards or away from the twisted states for $w=0.04$, $0.21$, $0.27$, and $0.42$. Solid orange, dashed blue, dot-dashed red, dotted green, and thick black curves represent systems with couplings given by only dyadic coupling, type a triadic coupling, type b triadic coupling, tetradic coupling, and mixed coupling, respectively. These winding numbers are also marked by vertical black lines in panels (a)--(c).}
    \label{fig1}
\end{figure*}

We consider the dynamics of $N$ coupled phase oscillators whose dynamics evolve according to
\begin{align}
\dot{\theta}_i&=\omega +\frac{K_1}{\langle k^{(1)}\rangle}\sum_{j=1}^NA_{ij}\sin(\theta_j-\theta_i)\nonumber\\
&+\frac{K_{2a}}{2\langle k^{(2)}\rangle}\sum_{j=1}^N\sum_{l=1}^NB_{ijl}\sin(\theta_j+\theta_l-2\theta_i)\nonumber\\
&+\frac{K_{2b}}{2\langle k^{(2)}\rangle}\sum_{j=1}^N\sum_{l=1}^NB_{ijl}\sin(2\theta_j-\theta_l-\theta_i)\nonumber\\
&+\frac{K_3}{6\langle k^{(3)}\rangle}\sum_{j=1}^N\sum_{l=1}^N\sum_{m=1}^NC_{ijlm}\sin(\theta_j+\theta_l-\theta_m-\theta_i),\label{eq:01}
\end{align}
where $\theta_i$ represents the phase of oscillator $i$, $\omega$ is the uniform natural frequency for all oscillators, coupling strengths $K_1$, $K_{2a}$, $K_{2b}$, and $K_3$ give the strengths of the four different kinds of viable interactions~\cite{Ashwin2016PhysD,Leon2019PRE}, and the adjacency matrix $A$ and tensors $B$ and $C$ encode the coupling structure. Here, we consider ring networks with interaction radius $r$, so entries of the adjacency matrix $A$ are $A_{ij}=1$ if $|i-j|\le r$ (modulo $N$) and otherwise $A_{ij}=0$. We also include triadic and tetradic interactions defined by the simplicial closure of the ring network, so $B_{ijl}=A_{ij}A_{jl}A_{li}$ and $C_{ijlm}=A_{ij}A_{jl}A_{lm}A_{mi}A_{il}A_{jm}$ (i.e., we ``fill in'' all three- and four-cliques). We note that coupling strengths are normalized by the appropriate multiple of the simplicial mean degrees $\langle k^{(1)}\rangle$, $\langle k^{(2)}\rangle$, and $\langle k^{(3)}\rangle$, each representing the mean number of unique 1-, 2-, and 3-simplexes (i.e., links, triangles, and tetrahedra) a node is part of. Moreover, without any loss of generality, we may set the natural frequency $\omega$ to zero by entering the rotating frame $\theta_i\mapsto \theta_i+\omega t$.

For the ring topology, the adjacency matrix $A$ and the adjacency tensors $B$ and $C$ are circulant, and the set of twisted states described by the vector $\bm{\theta}$ with entries $\theta_i=\theta_{\text{ref}}+qi2\pi/N$, parameterized by the winding number $q=1,\dots,N$ are fixed-point solutions of Eq.~(\ref{eq:01}). We note that each such twisted state is invariant under a constant shift, i.e., $\theta_i\mapsto\theta_i+\varphi$, so without any loss of generality we set the reference angle $\theta_{\text{ref}}$ to zero. The stability of a given twisted state $\bm{\theta}$ is determined by the eigenvalues of the Jacobian matrix $J(\bm{\theta})$, which is a linear combination of the Jacobians for the individual coupling terms, i.e.,
\begin{align*}
J(\bm{\theta})&=K_1 J_{1}(\bm{\theta}) + K_{2a}J_{2a}(\bm{\theta})+K_{2b}J_{2b}(\bm{\theta}) + K_3J_{3}(\bm{\theta}),
\end{align*}
where the individual Jacobians have off-diagonal entries 
\begin{align*}
    [J_{1}(\bm{\theta})]_{ij} &=  (1/\langle k^{(1)}\rangle)A_{ij}\cos(\theta_j-\theta_i),\\
    [J_{2a}(\bm{\theta})]_{ij} &= (1/2\langle k^{(2)}\rangle)\sum_{l=1}^N B_{ijl}\cos(\theta_j+\theta_l-2\theta_i),\\
    [J_{2b}(\bm{\theta})]_{ij} &= (1/2\langle k^{(2)}\rangle)\sum_{l=1}^N B_{ijl}\cos(2\theta_j-\theta_l-\theta_i),\\
    [J_{3}(\bm{\theta})]_{ij} &= (1/6\langle k^{(3)}\rangle)\sum_{l=1}^N B_{ijl}\cos(\theta_j+\theta_l-\theta_m-\theta_i),
\end{align*}
and diagonal entries balance the row-sums of the off-diagonal entries to attain a total row-sum of zero. For a system of $N$ oscillators, each twisted state we consider can then be parameterized by the normalized winding number, denoted here as $w=q/N\in[0,1]$, so that each twisted state may be written as $\bm{\theta}=\bm{\theta}(w)$, and the Jacobian can be similarly parameterized as $J(\bm{\theta}(w)):=J(w)$.

To explore the interplay between different kinds of higher-order coupling, we consider the dynamics given in Eq.~(\ref{eq:01}) on a ring network with interaction radius $r=3$ (which is the smallest $r$ that admits tetradic interactions). In Fig.~\ref{fig1} (a) we plot the dominant eigenvalue $\lambda_D$ of $J(w)$ that dictates stability as a function of the normalized winding number $w$. In particular, we plot $\lambda_D$ for the cases with only dyadic coupling, i.e., $K_1=1$ and all other coupling strengths are zero (orange diamonds), only the first type of tradic coupling (blue crosses), only the second type of tradic coupling (red squares), only tetradic coupling (green triangles), and finally, a combination of all couplings with $K_1=1/10$, $K_{2a}=1/2$, $K_{2b}=1/10$, and $K_3=3/10$ (black circles). These results (plotted using symbols) are directly computed from the Jacobian matrices, whereas the solid curves are given by the results obtained from considering the continuum limit, i.e., $N\to\infty$ given below [see Eqs.~(\ref{eq:05})--(\ref{eq:11})]. Given the complicated dependence of $\lambda_D$ on $w$, we plot in panel (b) a zoomed-in view near $\lambda_D\approx0$, where we can more clearly see when $\lambda_D<0$, i.e., for what $w$ the twisted state $\bm{\theta}(w)$ is stable. In panel (c), we illustrate the stability region for different twisted states for each case explicitly. We note that due to the translational symmetry described above, $J(\bm{\theta})$ always has one eigenvalue $\lambda=0$, associated with the constant eigenvector $\bm{v}\propto\bm{1}$, which we neglect. Moreover, there is a reflective symmetry about $w=0$ and $0.5$, which corresponds to odd symmetries in the twisted states, namely $\bm{\theta}(-w)=-\bm{\theta}(w)$ and $\bm{\theta}(0.5-w)=-\bm{\theta}(0.5+w)$. 

We observe that, while dyadic coupling stabilizes only states with winding number near zero, higher-order couplings additionally stabilize other states, e.g., islands of stability near $w\approx0.2$ and $0.5$. More importantly, as the winding number $w$ is varied, the stability of the twisted state changes for different coupling choices. For example, at $w=0.04$, the twisted state is stable for all the coupling schemes we consider. The stability of this twisted state can be seen explicitly in the dynamics of $\|\bm{\theta}(t)-\bm{\theta}^*\|$, where $\|\cdot\|$ is the 2-norm, representing the distance of a perturbed state from the reference twisted state $\bm{\theta}^*=\bm{\theta}^*(w)$, which we plot in panel (d) for all types of coupling schemes. In particular, we start with a perturbation of size $\|\bm{\theta}(t)-\bm{\theta}^*\|=10^{-1}$ and see all perturbations decay. Increasing $w$ to $0.21$, we see that both types of triadic coupling as well as mixed coupling yield stability, which we can similarly observe directly in panel (e). Further increasing $w$ to $0.27$, we see that only the ``2a'' type triadic coupling and mixed coupling yield stability, as shown in panel (f). Lastly, we find that for the winding number $w=0.42$, all four types of coupling on their own yield instability of $\bm{\theta}(w)$, but with mixed coupling this state becomes stable, as confirmed in panel (g).

\begin{figure}[t]
    \centering
    \includegraphics[width = 0.85\linewidth]{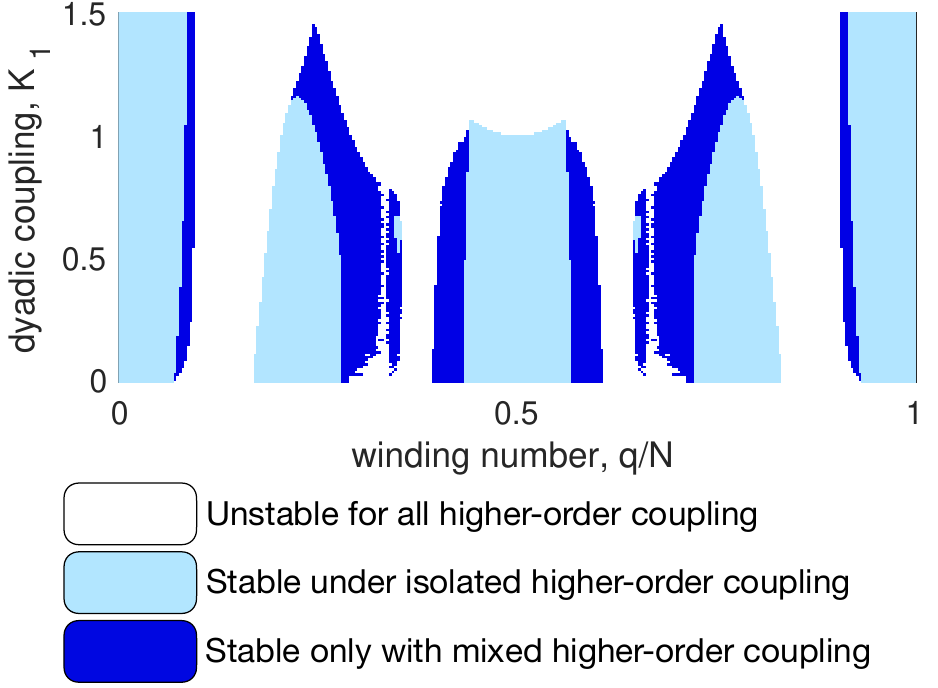}
    \caption{Stability regions of twisted states for different coupling combinations. For a network of size $N=200$ with interaction radius $r=3$, regions where the twisted state $\bm{\theta}(w)$ are not stable (white), stable under some combination of dyadic and only one type of higher-order coupling (light blue), and stable only under some combination that includes two or more types of higher-order coupling (dark blue) as a function of the winding number $w=q/N$ and dyadic coupling $K_1$.}
    \label{fig2}
\end{figure}

These results imply that the choices and combinations of higher-order couplings sensitively shape the emergent dynamics. In particular, studying each coupling in isolation is not enough to uncover the full range of collective dynamics, as a combination of higher-order couplings can give rise to dynamics that are not observed under any individual coupling. To demonstrate that this phenomenon is not limited to a small range of parameters, we consider the full range of combinations of the higher-order coupling strengths, $(K_{2a},K_{2b},K_3)$, with 
the strengths adding up to one. Using a ring network with $r=3$ and $N=200$, we vary the dyadic coupling strength $K_1$ and identify which twisted states $\bm{\theta}(w)$ are always unstable, which twisted states are stable when there is only one type of higher-order coupling present, i.e., with two of the three strengths in $(K_{2a},K_{2b},K_3)$ equal to zero, and which twisted states are stable only if there are at least two higher order coupling strengths in $(K_{2a},K_{2b},K_3)$ that are non-zero. Figure~\ref{fig2} shows the three categories above in white, light blue, and dark blue. We find that a wide range of parameters support a stable twisted state only with mixed higher-order couplings.

To better understand the stability under different coupling schemes, it is useful to note that, given the ring structure of the underlying network,  each individual Jacobian $J_{i}(\bm{\theta}(w))$ is circulant and can be written as (dropping the notation indicating dependence on $\bm{\theta}(w)$)
\begin{align}
J_i=(J_i^s)=\begin{bmatrix}
J_i^0 & J_i^1 & J_i^2 & \cdots & J_i^2 & J_i^1\\
J_i^1 & J_i^0 & J_i^1 & \cdots & J_i^3 & J_i^2\\
\vdots & \vdots & \vdots & \ddots & \vdots & \vdots \\
J_i^2 & J_i^3 & J_i^4 & \cdots & J_i^0 & J_i^1\\
J_i^1 & J_i^2 & J_i^3 & \cdots & J_i^1 & J_i^0
\end{bmatrix},\label{eq:03}
\end{align}
and the linear combination $J=K_1J_1+K_{2a}J_{2a}+K_{2b}J_{2b} + K_{3}J_3$ is also circulant. This implies that each Jacobian has the same set of eigenvectors $\{\bm{v}^p\}_{p=1}^N$ given by cosines, i.e., with the entries of the $p^{\text{th}}$ eigenvector given by $v_j^p=\cos(2\pi(p-1)j/N)$. Thus, rather than indexing eigenvectors by increasing or decreasing eigenvalues (as is often done), we index eigenvalues $\lambda_p$ by the associated eigenvectors $\bm{v}^p$. It follows that the $p^{\text{th}}$ eigenvalue of the full Jacobian $J$ is simply the linear combination of the $p^{\text{th}}$ eigenvalues of the individual Jacobians, i.e.
\begin{align}
\lambda_p = K_1\lambda_{1,p} + K_{2a}\lambda_{2a,p} + K_{2b}\lambda_{2b,p} + K_3 \lambda_{3,p}.\label{eq:04}
\end{align}
It is critical to point out, however, that the index $p$ for the dominant eigenvalue of the individual Jacobians need not be the same, and in general the dominant index of the full Jacobian need not coincide with any of the individual Jacobians. Thus, even if the dominant eigenvalues for $J_1$, $J_{2a}$, $J_{2b}$, and $J_3$ are all positive, the twisted state can still be stable under a combination of coupling types.

\begin{figure}[t]
    \centering
    \includegraphics[width = 0.95\linewidth]{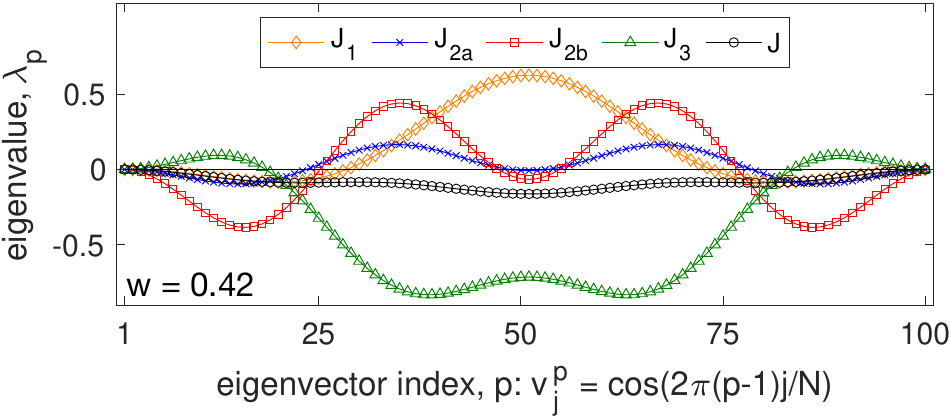}
    \caption{Balancing of the eigenvalue sprectra explains how mixed higher-order interactions can stabilize states that are unstable under any individual interaction. For a network of size $N=100$ with interaction radius $r=3$, the eigenvalues $\lambda_p$ of the Jacobians for the twisted state with $w=0.42$ as a function of the index $p$ defining the eigenvector $\bm{v}^p$ with indices $v_j^p=\cos[2\pi(p-1)j/N]$. Eigenvalues for Jacobians $J_1$, $J_{2a}$, $J_{2b}$, $J_3$, and the mixed coupling case $J=K_1J_1+K_{2a}J_{2a}+K_{2b}J_{2b}+K_3J_3$ are plotted in orange diamonds, blue crosses, red squares, green triangles, and black circles, respectively.}
    \label{fig3}
\end{figure}

We demonstrate this by plotting in Fig.~\ref{fig3} the eigenvalue $\lambda_p$ as a function of the normalized index $p/N$ corresponding to the eigenvector $\bm{v}^p$ for twisted states with $w=0.42$ on the network with interaction radius $r=3$ and size $N=100$, which is stable under mixed higher-order couplings with $K_1=1/10$, $K_{2a}=1/2$, $K_{2b}=1/10$, and $K_{3}=3/10$ but unstable under any individual coupling. We plot the eigenvalues $\lambda_p$ for the individual Jacobians $J_1$, $J_{2a}$, $J_{2b}$, and $J_3$ in orange diamonds, blue crosses, red squares, and green triangles, as well as the combined Jacobian in black circles. Importantly, while there are positive eigenvalues for each Jacobian $J_1$, $J_{2a}$, $J_{2b}$, and $J_3$, when combined the negative eigenvalues overcome the positive ones for all $p$, resulting all negative eigenvalues for $J$.

These results can be further developed analytically by writing down the entries $J_i^s$. 
For a given twisted state $\bm{\theta}(w)$ and coupling radius $r$, for $0<s\le r$,
\begin{align}
J_1^s &=\frac{1}{2r}\cos\left[2\pi(q/N)s\right],\label{eq:05}\\
J_{2a}^s &=\frac{2}{3r(r-1)}\sum_{\substack{k=s-r\\k\ne0,s}}^r\cos\left[2\pi(q/N)(s+k)\right],\label{eq:06}\\
J_{2b}^s &=\frac{1}{3r(r-1)}\sum_{\substack{k=s-r\\k\ne0,s}}^r\left\{2\cos\left[2\pi(q/N)(2s-k)\right]\right.\nonumber\\
&\hskip19ex\left.-\cos\left[2\pi(q/N)(s-2k)\right]\right\},\label{eq:07}\\
J_3^s &=\frac{1}{4r(r-1)(r-2)}\nonumber\\
&\times\sum_{\substack{k=-r\\k\ne0\\0<|k-s|\le r}}^r\sum_{\substack{l=-r\\l\ne0\\0<|l-s|\le r\\0<|l-k|\le r}}^r\left\{2\cos\left[2\pi(q/N)(s+k-l)\right]\right.\nonumber\\
&\hskip18ex\left.-\cos\left[2\pi(q/N)(s-k-l)\right]\right\}.\label{eq:08}
\end{align}
Using the fact that the eigenvectors $\{\bm{v}^p\}_{p=1}^N$ of each Jacobian $J_i$ is given by a cosine, i.e., $v_j^p=\cos(2\pi(p-1)j/N)$, the eigenvalues of the Jacobian $J_i$ can be written exactly as
\begin{align}
\lambda_p=\sum_{s=0}^{N-1}J_i^s e^{2\pi i(p-1)s/N},\label{eq:09}
\end{align}
where we assume that $J_i^s=J_i^{N-s}$ for $s>\lfloor N/2\rfloor$.
Equation~(\ref{eq:09}) can be used directly to calculate both the eigenvalues of any individual Jacobian $J_i$ or the full Jacobian $J=K_1J_1+K_{2a}J_{2a}+K_{2b}J_{2b}+K_3J_3$.

\begin{figure}[t]
    \centering
    \includegraphics[width = 1.0\linewidth]{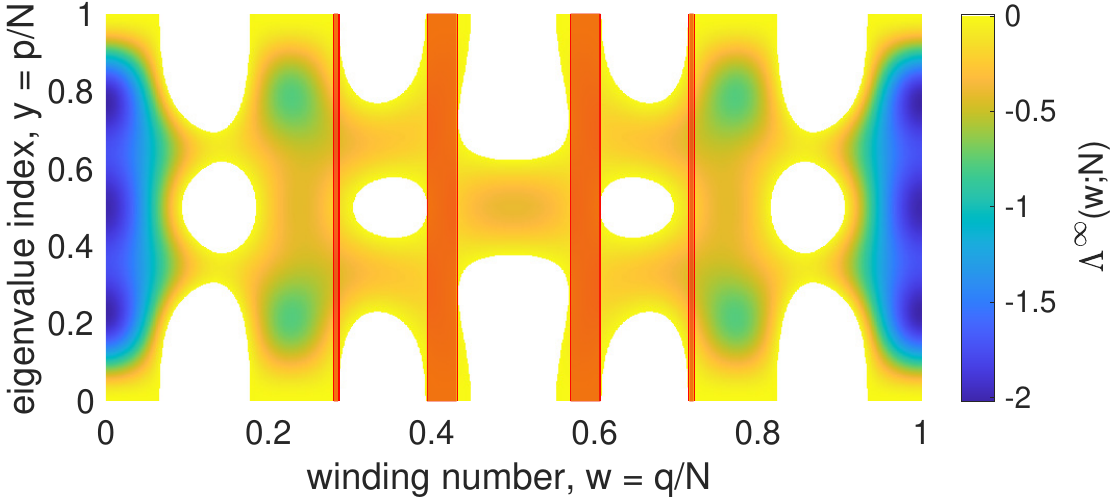}
    \caption{Heat map for the continuum limit eigenvalue function $\lambda^\infty(w,y)$ for the case $r=3$ with coupling $K_1=1/10$, $K_{2a}=1/2$, $K_{2b}=1/10$, and $K_3=3/10$. 
    {Regions where $\lambda^\infty(w,y)>0$ are colored white, i.e., the $y^\text{th}$ mode of the $w^\text{th}$ twisted state is unstable. The $w^\text{th}$ twisted state is stable when $\lambda^\infty(w,\cdot)<0$.} 
    The region $\mathcal{W}/\left(\mathcal{W}_1\cup\mathcal{W}_{2a}\cup\mathcal{W}_{2b}\cup\mathcal{W}_3\right)$, where only a mixture of higher-order coupling yields stability, is indicated with the red vertical bars.}
    \label{fig4}
\end{figure}

These results can be taken one step further by considering the continuum limit $N\to\infty$. For this purpose, it is useful to simplify our notation for the Jacobian matrices evaluated at the twisted state $\bm{\theta}(w)$, i.e., $J_i(w)=J_i(\bm{\theta}(w))$, where we recall that $w=q/N$ is the normalized winding number. Moreover, we let $y=p/N$ for $p=1,\dots,N$ represent the normalized eigenvector and eigenvalue index. We then define the function
\begin{align}
\lambda_i^\infty(w,y)=\sum_{s=-r}^rJ_i^s(w)\cos(2\pi s y),\label{eq:10}
\end{align}
which represents the $y^{\text{th}}$ eigenvalue corresponding to the $w^{\text{th}}$ twisted state for the Jacobian $J_i$ in the continuum limit. For predictive power of stability for a network of size $N$, we then define the dominant non-trivial eigenvalue as the maximum of Eq.~(\ref{eq:10}) over indices $y=1/N,2/N,\dots,(N-1)/N$, i.e.,
\begin{align}
\Lambda_i^\infty(w;N)=\max_{y\in\{1/N,2/N,\dots,(N-1)/N\}}\lambda_i^\infty(w,y),\label{eq:11}
\end{align}
where the winding number $w$ is also constrained to take on values $0,1/N,\dots,(N-1)/N$. This continuum limit formulation of the dominant eigenvalue gives the solid curves plotted in Fig.~\ref{fig1}(a) and (b), matching well with numerics. Moreover, the predicted region of stability is the set $\mathcal{W}_i=\{w\in[0,1]|\Lambda_i^\infty(w;N)<0\}$, so that the set of twisted states $w$ for which $\bm{\theta}(w)$ is stable only when all couplings are present (i.e., given $K_1$, $K_{2a}$, $K_{2b}$, and $K_3$) is
\begin{align}
\mathcal{W}/\left(\mathcal{W}_1\cup\mathcal{W}_{2a}\cup\mathcal{W}_{2b}\cup\mathcal{W}_3\right).\label{eq:12}
\end{align}
In Fig.~\ref{fig4} we plot the continuum-limit eigenvalue function $\lambda^\infty(w,y)$ for the case of combined coupling with $K_1=1/10$, $K_{2a}=1/2$, $K_{2b}=1/10$, and $K_3=3/10$, with positive values colored white. We also shade in red the region $\mathcal{W}/\left(\mathcal{W}_1\cup\mathcal{W}_{2a}\cup\mathcal{W}_{2b}\cup\mathcal{W}_3\right)$, which we find to be approximately the set $[0.2788,0.2856]\cup[0.3944,0.4308]\cup[0.5692,0.6056]\cup[0.7144,0.7212]$.

In this Letter, we have demonstrated that when different higher-order coupling functions are simultaneously considered, they can stabilize new dynamical states that are not stable under any individual higher-order interaction. 
We support this observation by presenting an exact description of the Jacobian eigenvalue spectrum for all twisted states on ring simplicial complexes.
While coupled oscillators on ring simplicial complexes is a fertile ground for understanding this phenomena, future works have the opportunities to explore how new states might emerge from different coupling topologies and with heterogeneous local dynamics (e.g., heterogeneous natural frequencies). 

\acknowledgements
PSS acknowledges support from NSF grant MCB 2126177. 
FB acknowledges support from the Austrian Science Fund (FWF) through project 10.55776/PAT1052824 and project 10.55776/PAT1652425. 
ML is a Postdoctoral Researcher of the Fonds de la Recherche Scientifique–FNRS. 
MSM acknowledges support from NSF grant DMS 2406942 and use of the ELSA high performance computing cluster at The College of New Jersey for conducting the research reported in this paper which was funded in part by NSF grants OAC 1826915 and OAC 2320244.

All data corresponding to the findings in this manuscript is available upon reasonable request from the corresponding author.

%




\bibliography{bib}

\end{document}